\documentclass{SciPost}

\hypersetup{
    colorlinks,
    linkcolor={red!50!black},
    citecolor={blue!50!black},
    urlcolor={blue!80!black}
}

\usepackage[bitstream-charter]{mathdesign}
\DeclareSymbolFont{usualmathcal}{OMS}{cmsy}{m}{n}
\DeclareSymbolFontAlphabet{\mathcal}{usualmathcal}

\fancypagestyle{SPstyle}{
\fancyhf{}
\lhead{\colorbox{scipostdeepblue}{\bf \color{white} ~SciPost Physics Proceedings }}
\rhead{{\bf \color{scipostdeepblue} ~Submission }}

\fancyfoot[C]{\textbf{\thepage}}
}

\begin{document}

\pagestyle{SPstyle}

\begin{center}{\Large \textbf{\color{scipostdeepblue}{
Applications of Machine Learning \\ in Constraining Multi-Scalar Models\\
}}}\end{center}

\begin{center}\textbf{
Darius~Jur\v{c}iukonis\textsuperscript{1$\star$}
}\end{center}

\begin{center}
{\bf 1} Vilnius University, Institute of Theoretical Physics and Astronomy
\\[\baselineskip]
$\star$ \href{mailto:darius.jurciukonis@tfai.vu.lt}{\small darius.jurciukonis@tfai.vu.lt}
\end{center}

\definecolor{palegray}{gray}{0.95}
\begin{center}
\colorbox{palegray}{
  \begin{tabular}{rr}
  \begin{minipage}{0.37\textwidth}
    \includegraphics[width=60mm]{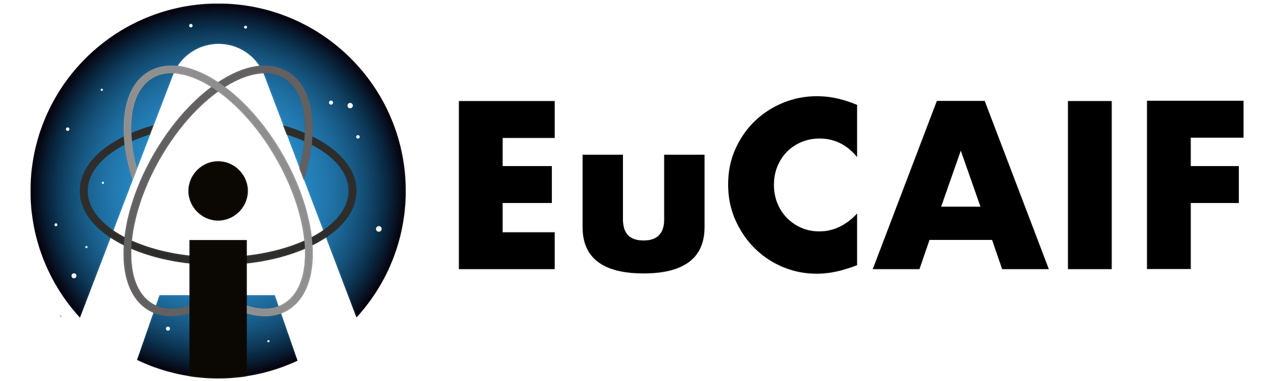}
  \end{minipage}
  &
  \begin{minipage}{0.5\textwidth}
    \vspace{5pt}
    \vspace{0.5\baselineskip} 
    \begin{center} \hspace{5pt}
    {\it The 2nd European AI for Fundamental \\Physics Conference (EuCAIFCon2025)} \\
    {\it Cagliari, Sardinia, 16-20 June 2025
    }
    \vspace{0.5\baselineskip} 
    \vspace{5pt}
    \end{center}
    
  \end{minipage}
\end{tabular}
}
\end{center}

\section*{\color{scipostdeepblue}{Abstract}}
\textbf{\boldmath{%
Machine learning techniques are used to predict theoretical constraints such as unitarity and boundedness from below in extensions of the Standard Model.
This approach has proven effective for models incorporating additional $SU(2)$ scalar multiplets, in particular the quadruplet and sixplet cases.
High predictive performance is achieved through the use of suitable neural network architectures and well-prepared training datasets.
Moreover, machine learning provides a substantial computational advantage, enabling significantly faster evaluations compared to scalar potential minimization.
}}

\vspace{\baselineskip}

\noindent\textcolor{white!90!black}{%
\fbox{\parbox{0.975\linewidth}{%
\textcolor{white!40!black}{\begin{tabular}{lr}%
  \begin{minipage}{0.6\textwidth}%
    {\small Copyright attribution to authors. \newline
    This work is a submission to SciPost Phys. Proc. \newline
    License information to appear upon publication. \newline
    Publication information to appear upon publication.}
  \end{minipage} & \begin{minipage}{0.4\textwidth}
    {\small Received Date \newline Accepted Date \newline Published Date}%
  \end{minipage}
\end{tabular}}
}}
}




\section{Introduction}
\label{sec:intro}

Theoretical consistency in high-energy physics models is ensured by constraints such as unitarity (UNI) and bounded-from-below (BFB) conditions, which play a central role in establishing both physical viability and vacuum stability.
Unitarity imposes restrictions on the behavior of scattering amplitudes, guaranteeing that they remain finite and well-defined.
In contrast, the BFB condition ensures the stability of the vacuum by requiring that the scalar potential be bounded from below, thereby preventing the vacuum from decaying into states of arbitrarily lower energy.

Unitarity conditions can be evaluated analytically, through the computation of the eigenvalues of scattering matrices.
Deriving analytical vacuum stability conditions is generally a challenging task and feasible only for relatively simple models.
For more complex scenarios, one must rely on numerical minimization of the scalar potential, a process that is often computationally expensive.

In this study, two extensions of the Standard Model (SM) with additional $SU(2)$ multiplets, specifically the 4-plet and 6-plet cases, are investigated.
The analytical UNI conditions derived in Ref.~\cite{Jurciukonis:2024cdy} together with the BFB conditions recently presented in Ref.~\cite{milagre} provide a framework to assess the reliability of machine learning (ML) approaches for this class of models.
The results are compared with those of the general two-Higgs-doublet model (G2HDM), for which the BFB conditions can be determined using a simple and precise algorithm~\cite{2hdm,ML}.
Machine learning approaches are finding an increasing number of applications in theoretical physics (see, e.g., Refs.~\cite{Binjonaid:2024jpm,Hammad:2024tzz}); 
therefore, the method discussed in this paper can also be employed for efficient scanning of parameter spaces alongside traditional techniques.

\section{Strategy}
\label{sec:strategy}

\begin{figure}[ht]
\begin{center}
\includegraphics[width=0.9\textwidth]{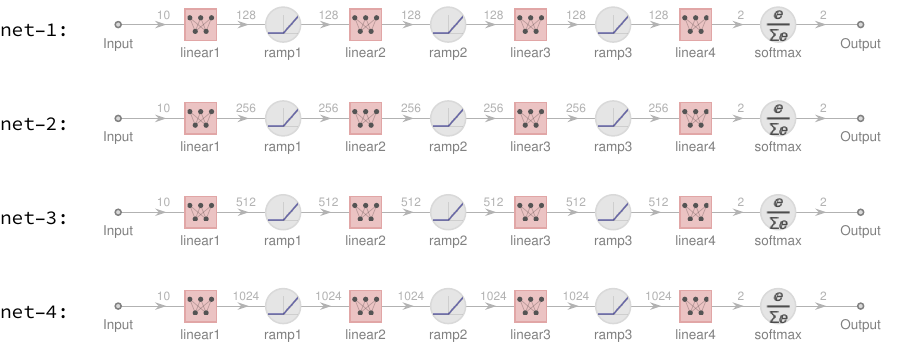}
\end{center}
\vspace{-4mm}
\caption{\small{The architecture of the networks used in the analysis of the G2HDM is shown as an example. The ten input parameters correspond to the G2HDM potential couplings, with their real and imaginary parts treated separately.}}
\label{picture1}
\end{figure}

Adopting the computational strategy proposed in Ref.~\cite{ML}, we train neural networks to predict the UNI and BFB constraints simultaneously.
For that, we employ four fully connected neural networks, each consisting of eight layers as displayed in Fig.~\ref{picture1}.
The first network, net-1, comprises 128 neurons and is trained on raw data supplemented with an appropriate number of true samples.
Net-1 is then used to generate refined training data for net-2, which has 256 neurons.
Subsequently, net-3 and net-4 are trained on the same dataset as net-2 but with larger architectures of 512 and 1024 neurons, respectively.
The outputs of net-3 and net-4 are further used to filter the predictions obtained from net-2.

The first two networks provide faster inference but lower accuracy when operating on raw data, whereas net-3 and net-4 achieve higher accuracy at the cost of increased computational time. 
The number of layers and neurons in the networks was selected to optimize the tune-up between accuracy and computational efficiency. 
Consequently, this combination of simpler and more complex networks enables high overall accuracy while maintaining efficient inference speed.

For this study, the neural networks were trained on an NVIDIA GeForce RTX 4090 GPU with 24\,GB of VRAM. Inference and additional computations were performed on a desktop computer equipped with an Intel Core i9-13900K CPU (24 cores) and 128 GB of RAM. All cases under comparison were evaluated under identical conditions, with calculations executed in parallel across all available CPU cores.

Training was performed using datasets containing $10^6$--$10^7$ samples, over 500 training epochs with a batch size of $2^{14}$ samples. The training time ranged from 10 to 60 minutes, depending on the network size, and required up to 2\,GB of GPU memory.

Since Wolfram Mathematica is widely used in the high-energy physics (HEP) community, all ML and numerical analyses were conducted within {\tt Mathematica} to ensure compatibility with existing HEP packages.
The functions and networks, together with computational examples 
are available at 
\href{https://github.com/jurciukonis/ML-for-multiples}{{\tt https://github.com/jurciukonis/ML-for-multiples}}.

\section{Results}
\label{sec:results}

Randomly generated sets of potential parameters (raw data) contain only a small fraction of samples that satisfy the UNI conditions, the BFB conditions, or both, as shown in Table~\ref{table_percentage}.
Consequently, the raw data must be augmented with valid samples to ensure that the training dataset is sufficiently informative for effective network training.

\begin{table}[ht]
  \begin{center}
    \begin{tabular}{|c|c|c|c|c|}
    \hline
     Model & \hspace*{3.mm} UNI \hspace*{3.mm} & \hspace*{3.mm} BFB-I \hspace*{3.mm} & \hspace*{3.mm} BFB-II \hspace*{3.mm} &
     UNI+BFB \\ \hline \hline
     G2HDM \,(10) & 1.02\% & 23.3\% & 3.25\% & 0.033\% \\
     SM + 4plet \,(5) & 12.8\% & 30.0\% & 17.3\% & 2.2\% \\
     SM + 6plet \,(6) & 5.4\% & 22.2\% & 6.07\% & 0.33\% \\
     \hline
    \end{tabular}
  \end{center}
  \vspace*{-4mm} 
\captionof{table}{Percentage of true samples in the models with randomly generated parameters.
The number of model parameters to be generated is indicated in the brackets in the first column.
The “BFB-I” column reports the fraction of true samples in the raw datasets, while the “BFB-II” column shows the corresponding fraction after refinement using the UNI conditions.\protect\footnotemark \, 
The last column reports the percentage of samples that simultaneously satisfy both UNI and BFB conditions.
    \label{table_percentage}}
\end{table}
\footnotetext{Since the UNI conditions are more restrictive than the BFB conditions and constrain the parameter space, some samples satisfying the BFB criteria are discarded in the “BFB-II” column.}

Although the models under consideration admit analytical BFB conditions, they can also be evaluated through potential minimization, allowing direct comparison with computations performed using neural networks.
The prediction of UNI and BFB conditions with ML can be accelerated by more than a factor of 400 in the case of the SM + 4plet model and by more than a factor of 600 in the case of the SM + 6plet model, relative to calculations based on potential minimization, as shown in Table~\ref{table_times}.
These predictions can subsequently be validated via the minimization procedure, while still maintaining computationally manageable times, as indicated in the last column of Table~\ref{table_times}.

\begin{table}[ht]
  \begin{center}
    \begin{tabular}{|c|c|c|c|c|c|}
    \hline
     Model & \hspace*{.mm} UNI+min. \hspace*{0.mm} & neural nets & \hspace*{0.mm} neural nets+min. \hspace*{0.mm} &
     \hspace*{0.mm} ratio-I \hspace*{0.mm} & \hspace*{0.mm} ratio-II \hspace*{0.mm} \\ \hline \hline
     G2HDM & 131 & 1.7 & 35 & 77 & 3.7 \\
     SM + 4plet & 53 & 0.12 & 19 & 441 & 2.7 \\
     SM + 6plet & 496 & 0.76 & 45 & 653 & 11 \\
     \hline
    \end{tabular}
  \end{center}
  \vspace*{-4mm} 
\captionof{table}{Computation times (in seconds) required to identify \emph{1000 true samples}.
The second column presents the computation times obtained using both UNI conditions and global minimization.
The third column shows the corresponding times when employing neural networks alone, while the fourth column provides the times obtained by combining neural network predictions with subsequent verification through global minimization.
    \label{table_times}}
\end{table}

Since the raw data consist predominantly of false samples, even net-1 achieves very high accuracy in data classification.
Therefore, to enable a more meaningful comparison, we focus on the fraction of true samples correctly classified by the networks (i.e. true positives), as verified using the analytical UNI + BFB conditions.~\footnote{Alternatively, the false positive rate can be estimated using the relation 
\begin{equation}
\mathrm{FPR} = \frac{100 p_1 - p_1 p_2}{100 \left(p_1 + p_2 \right) - 2p_1 p_2},
\end{equation}
where $p_1$ denotes the percentages reported in the last column of Table~\ref{table_percentage}, and $p_2$ denotes the corresponding percentages from Table~\ref{table_percentage2} for each model.
For example, for the network combination net-2 -- net-4 applied to the SM + 6plet model, we obtain $\mathrm{FPR} = 1.7 \times 10^{-5}$.
}

For models with additional multiplets, the accuracy of net-1 remains relatively high; however, it can be substantially improved by employing more complex networks or network combinations, as shown in Table~\ref{table_percentage2}.

\begin{table}[ht]
  \begin{center}
    \begin{tabular}{|c|c|c|c|c|}
    \hline
     Model & \hspace*{3.mm} net-1 \hspace*{3.mm} & \hspace*{3.mm} net-2 \hspace*{3.mm} & \hspace*{3.mm} net-3,4 \hspace*{3.mm} &
     net-2 -- net-4 \\ \hline \hline
     G2HDM & 52--55\% & 96--97\% & 97--98\% & $> 99.0\%$ \\
     SM + 4plet & 98--99\% & 98.8--99.1\% & $\sim 99.2\%$ & $> 99.5\%$ \\
     SM + 6plet & 92--94\% & 98.8--99.2\% & $\sim 99.0\%$ & $> 99.5\%$ \\
     \hline
    \end{tabular}
  \end{center}
  \vspace*{-4mm} 
\captionof{table}{Percentage of true samples in the predicted results, validated against exact analytical UNI and BFB conditions.
    \label{table_percentage2}}
\end{table}

\section{Conclusion}

In this work, we investigated the reliability of machine learning methods for extensions of the Standard Model with additional $SU(2)$ multiplets.
We demonstrated that ML techniques can efficiently predict UNI and BFB constraints in models with added 4-plet and 6-plet.
This approach achieves high predictive accuracy while significantly reducing computation time compared to global minimization of the scalar potential.
Although the models considered here are relatively simple, their parameter spaces become substantially more intricate when, for example, specific hypercharge assignments are introduced, as discussed in Refs.~\cite{4plet-1,4plet-2}.
In such cases, neural networks could provide an effective tool for constraining the parameter space.
Our results further indicate that this approach can be extended to more complex scenarios, such as the SM with additional 7-plet or 8-plet, for which analytical BFB conditions are not yet available.

\section*{Acknowledgements}
This work has received funding from the Research Council of Lithuania under Contract No. S-CERN-24-2.




\end{document}